\documentclass{article}

\usepackage[preprint]{mlsys2025}

\usepackage{microtype}
\usepackage{booktabs} 
\usepackage{colortbl}
\usepackage{tikz}
\usepackage{enumitem}
\usepackage{verbatim}
\usepackage{float}
\usepackage{caption}
\usepackage{amsmath,amsthm,amsfonts,amssymb,amscd,bm,amsbsy}
\usepackage{balance}
\usepackage[TABBOTCAP]{subfigure}
\usepackage{fancyhdr}
\usepackage{pifont}
\usepackage{mathrsfs}
\usepackage{listings}
\usepackage{textcomp}
\usepackage{xcolor}
\usepackage{hyperref}
\usepackage{ragged2e}
\usepackage[utf8]{inputenc}
\usepackage{graphicx}
\usepackage{lipsum}
\usepackage{indentfirst}
\usepackage{cleveref}
\usepackage{tabularx}
\usepackage{balance}
\usepackage{wrapfig}
\usepackage{multirow}
\usepackage{capt-of}
\usepackage[many]{tcolorbox}
\usepackage{placeins}
\usepackage{stfloats}
\makeatletter 
\usepackage{algorithm}
\definecolor{gjr}{HTML}{FF0000}

\newcommand{\ourname}{RTP-LLM}

\mlsystitlerunning{\ourname{}: Building an Evolvable and Reliable LLM Inference Engine at Alibaba Scale}

\begin{document}

\twocolumn[
\mlsystitle{\ourname{}: Building an Evolvable and Reliable LLM Inference Engine at Alibaba Scale}

\mlsyssetsymbol{equal}{*}

\begin{mlsysauthorlist}
\mlsysauthor{Jiarui Guo}{equal,pku,ali}
\mlsysauthor{Zhi Zhang}{equal,ali}
\mlsysauthor{Boyu Tan}{equal,ali,zju}
\mlsysauthor{Zongwei Lv}{pku}
\mlsysauthor{Haobo Sun}{ali,zju}
\mlsysauthor{Na He}{ali}
\mlsysauthor{Ziqing Wang}{ali}
\mlsysauthor{Xinfei Shi}{ali}
\mlsysauthor{Wei Zhang}{ali}
\mlsysauthor{Xi Yang}{ali}
\mlsysauthor{Haosheng Hu}{ali}
\mlsysauthor{Wending Bao}{ali}
\mlsysauthor{Zetao Hu}{ali}
\mlsysauthor{Dong Cao}{ali}
\mlsysauthor{Yong Li}{ali}
\mlsysauthor{Lining Wang}{ali}
\mlsysauthor{Chengjun Yang}{ali}
\mlsysauthor{Tianpeng Wang}{ali}
\mlsysauthor{Huanbing Lu}{ali}
\mlsysauthor{Yidan Wang}{ali}
\mlsysauthor{Hengyi Bao}{ali}
\mlsysauthor{Yaxin Yu}{ali}
\mlsysauthor{Hanyun Guo}{ali}
\mlsysauthor{Huidong Zhang}{ali}
\mlsysauthor{Haiyu Zhou}{ali}
\mlsysauthor{Zefeng Liu}{ali}
\mlsysauthor{Weixiang Huang}{ali}
\mlsysauthor{Ziyi Wang}{ali}
\mlsysauthor{Chenyi Song}{ali}
\mlsysauthor{Ruixuan Zhang}{ali}
\mlsysauthor{Zhengkai Ao}{ali}
\mlsysauthor{Haocheng Luo}{ali}
\mlsysauthor{Zhichen Xue}{ali}
\mlsysauthor{Jialin Zhang}{ali}
\mlsysauthor{Chi Zhang}{ali}
\mlsysauthor{Lingjun Zhu}{ali}
\mlsysauthor{Silu Zhou}{ali}
\mlsysauthor{Zongyuan Wu}{ali}
\mlsysauthor{Yinghao Yu}{ali}
\mlsysauthor{Yi Zheng}{ali}
\mlsysauthor{Miao Sun}{ali}
\mlsysauthor{Yuxing Yuan}{ali}
\mlsysauthor{Juncheng Yin}{ali}
\mlsysauthor{Guoding Li}{ali}
\mlsysauthor{Nan Wang}{ali}
\mlsysauthor{Xiyu Wang}{ali}
\mlsysauthor{Lin Yang}{ali}
\mlsysauthor{Zechao Zhang}{ali}
\mlsysauthor{Jianning Zhang}{ali}
\mlsysauthor{Bo Cai}{ali}
\mlsysauthor{Xiaolin Wang}{pku}
\mlsysauthor{Tao Lan}{ali}
\mlsysauthor{Tong Yang}{pku}
\mlsysauthor{Kan Liu}{ali}
\mlsysauthor{Lu Chen}{zju}

\end{mlsysauthorlist}

\mlsysaffiliation{pku}{Peking University}
\mlsysaffiliation{ali}{Alibaba Group}
\mlsysaffiliation{zju}{Zhejiang University}
\mlsyscorrespondingauthor{Tong Yang}{yangtong@pku.edu.cn}
\mlsyscorrespondingauthor{Kan Liu}{liukan.lk@alibaba-inc.com}
\mlsyscorrespondingauthor{Lu Chen}{luchen@zju.edu.cn}

\vskip 0.3in

\begin{abstract}
Large language models and inference techniques evolve rapidly, making model enablement a recurring production task rather than a one-time integration effort. 
At Alibaba, a shared inference infrastructure must keep pace with new models and serving capabilities while remaining reliable across heterogeneous production environments.
To reconcile rapid evolution with production reliability, we present \ourname{}, an open-source production inference engine developed and deployed at Alibaba. 
Its architecture separates model-specific execution from shared request processing, state management, and resource coordination.
Explicit interfaces localize compatible model changes while allowing models and deployment configurations to reuse common serving mechanisms.
\ourname{} has evolved from an exploratory system into a shared production engine that supports $\sim$100 models and processes hundreds of trillions of tokens per day.
We describe how its architecture and production practices accommodate model evolution and contain the impact of failures, and share lessons for building evolvable and reliable inference engines at enterprise scale.
\end{abstract}

]

\printAffiliationsAndNotice{\mlsysEqualContribution}

\section{Introduction}

In recent years, large language models (LLMs) have advanced at a remarkable pace \cite{gemma,gpt4}. 
Research institutions and technology companies continue to release open models spanning diverse parameter scales, architectures, and modalities \cite{mistral7B,deepseekv2,qwen2-vl}. 
Even within a single model family, new checkpoints and variants are released frequently \cite{qwen2.5}. 
In parallel, techniques for LLM inference continue to evolve, including low-bit quantization \cite{xiao2023smoothquant,lin2024awq}, speculative decoding \cite{leviathan2023fast}, diverse parallelization strategies \cite{narayanan2021efficient, rajbhandari2022deepspeed, liu2024ringattention} and prefill-decode disaggregation \cite{zhong2024distserve, patel2024splitwise}. 
Model enablement must therefore accommodate changes in both model computation and the techniques used to execute it.

However, a publicly released model checkpoint does not by itself constitute a production-ready service; it must be integrated into an inference engine and qualified under production performance and reliability requirements \cite{yu2022orca,vllm}.
At Alibaba, this task is amplified by the breadth and heterogeneity of its businesses. 
A shared inference infrastructure serves domestic and international commerce, logistics, digital mapping, and local consumer services, including Alibaba Cloud, Taobao, Tmall, Xianyu, Cainiao, Amap, Ele.me, AliExpress, and Lazada. 
These businesses differ in their models, hardware platforms, deployment configurations, workload characteristics, and service-level objectives. 
Model adaptation is therefore not an isolated compatibility exercise for a single application, but a recurring production requirement through which new model capabilities are made available across the company.

Supporting this range of workloads through a common inference foundation imposes two competing requirements. 
New models and serving capabilities must be incorporated rapidly, while existing services must retain correctness, predictable performance, and availability through frequent changes.
Insufficiently isolated modifications can disrupt unrelated deployments, whereas interfaces made too rigid in pursuit of stability can make each new model or feature costly to integrate.
\textit{The central challenge is therefore not merely to support many models or maximize peak performance, but to evolve continuously without compromising the reliability of existing services.}

Existing inference frameworks have made substantial advances in performance, model coverage, and serving functionality \cite{vllm,sglang,tensorrt-llm}. 
However, adopting them for Alibaba's production environment still requires substantial local validation and integration.
First, upstream validation cannot cover every combination of workloads, hardware, and operating conditions encountered at Alibaba.
In our experience, some correctness, cache-consistency, and overload-related bugs emerge only under sustained traffic, heterogeneous hardware, or fleet-scale deployment, and may remain invisible in functional tests and small-scale benchmarks. 
Also, the broad feature and component surface of a general-purpose engine may further expand the code that must be integrated, validated, and maintained, even when only a subset is used internally.
Second, a deeply customized fork can address local requirements, but the necessary changes may span scheduling, cache management, hardware backends, and other shared components.
As the upstream project evolves, these modifications must be repeatedly reconciled with upstream changes, creating persistent integration effort, regression risk, and maintenance cost. 
These trade-offs motivated our decision to own the engine's architectural evolution and production validation.

To realize this objective, we present \ourname{}, an open-source production inference engine developed and deployed at Alibaba. 
Its central design principle is to separate rapidly changing model execution from relatively stable serving control. 
Explicit interfaces keep compatible model-specific changes within the execution layer while allowing serving mechanisms to be reused across models and deployment configurations.
Our major contributions are summarized as follows:

\begin{itemize}[leftmargin=1em]
    \item We design, implement, and open-source \ourname{} \cite{rtp-llm}, a production inference engine that separates model execution from stateful serving control.\footnote{
        The source code of \ourname{} is available at \url{https://github.com/alibaba/rtp-llm}.
    }
    We present its extensible model-execution layer and shared runtime mechanisms for request processing, state management, and resource coordination.

    \item We report the large-scale deployment and continuous evolution of \ourname{} at Alibaba.
    It serves multiple business domains and has supported successive generations of models, accelerator platforms, and serving capabilities.

    \item We share experiences and lessons from the long-term development and operation of \ourname{}.
    We analyze critical design trade-offs and provide practical guidelines for building evolvable and reliable production inference engines.
\end{itemize}

\section{Background}

\subsection{From Model Release to Production Service}
\label{sec:model-readiness}

A model release typically provides a checkpoint, model configuration, tokenizer, and reference implementation. 
These artifacts specify the model computation but not how it should operate as an online service. 
An inference engine \cite{vllm, sglang, tensorrt-llm,deepinfer2026} bridges this gap by executing the model, coordinating parallel computation, scheduling and batching requests, managing KV caches and accelerator resources, and returning outputs under performance and reliability requirements.

Based on our production experience at Alibaba, we organize model enablement into six readiness states. 
A model progresses from \emph{available}, to \emph{correctness-ready} when its outputs satisfy defined comparisons against a reference implementation, and then to \emph{performance-ready} when it satisfies workload-specific latency, throughput, and resource-efficiency targets.
It becomes \emph{integration-ready} after connecting to the surrounding service infrastructure, \emph{canary-ready} when it can safely receive restricted production traffic, and finally \emph{production-ready} when it can sustain production traffic under the required service-level objectives. 
Although validation activities may overlap and failures may trigger iteration, these states represent progressively stronger readiness gates.
Throughout this paper, model enablement refers to this complete transition, rather than merely the successful execution of a model checkpoint.

\subsection{Evolution towards \ourname{}}

The current architecture of \ourname{} emerged through three stages, each prompted by assumptions that no longer held as models and production requirements evolved. 

\textbf{Stage I: FasterTransformer-based exploration.} 
The earliest system was developed between March 2023 and June 2024.
Throughout this stage, the models targeted by the system were predominantly dense, decoder-only Transformers with largely similar attention structures, differing mainly in components such as positional encoding and feed-forward networks \cite{llama,qwen}.
It used FasterTransformer's C++/CUDA model implementations as its execution backend \cite{fastertransformers}, while a Python service layer handled tokenization, request scheduling, streaming responses, health checks, and request-level KV-cache lifecycle management.
This design was sufficient for early exploration, but it became inadequate as the system expanded. 
On some early workloads, the Python service layer incurred 10--20\,ms of host-side overhead per model invocation.
This overhead became significant when GPU execution was short.
The model implementation inherited from FasterTransformer was also difficult to extend to new accelerator platforms, parallel configurations, and model structures.

\textbf{Stage II: C++-first integration.} 
Starting in June 2024, \ourname{} moved batching, scheduling, and KV-cache management into C++ and replaced the inherited FasterTransformer backend with a C++-first model-execution path. 
Compared with the earlier Python service layer, this design substantially reduced host-side serving overhead.
It also introduced a device abstraction through which the model-forward path selected backend-specific operators, enabling a common runtime to support heterogeneous accelerators.
During its subsequent evolution, it incorporated optimized attention kernels, MoE parallelism, and prefill--decode disaggregation \cite{fa1, fa2, deepseekv3, deepep2025, zhong2024distserve}.
However, its C++-first organization became increasingly costly as new models and operators were primarily released as Python/PyTorch implementations in the Hugging Face ecosystem \cite{paszke2019pytorch, wolf2020transformers}.
Repeatedly translating model-specific semantics into C++ made adaptation time difficult to predict, complicated the reuse of native operators, and increased maintenance effort, eventually motivating another architectural revision.

\textbf{Stage III: Decoupled model execution and serving mechanisms.} 
In June 2025, the current \ourname{} architecture moved model execution and model-specific semantics to Python and PyTorch, while retaining batching, scheduling, KV-cache management, and resource coordination in a shared high-performance C++ runtime. 
CUDA Graphs capture and replay repeated GPU execution \cite{cudagraph}, reducing host-side launch overhead while model logic remains in Python/PyTorch.
This separation allows new model structures and operators to be expressed closer to their reference implementations, while reusing common serving mechanisms across models and deployments \cite{glm5, deepseekv4, kimik3}. 
The boundary is not immutable: model changes that alter shared state or resource requirements may still require coordinated runtime changes. 
Nevertheless, \ourname{} has continued to incorporate new models and serving capabilities without another fundamental architectural redesign.

\section{RTP-LLM System Overview}
\label{sec:architecture}

\begin{figure}[t]
    \centering
    \includegraphics[width=0.95\linewidth,trim=0 15bp 0 12bp,clip]{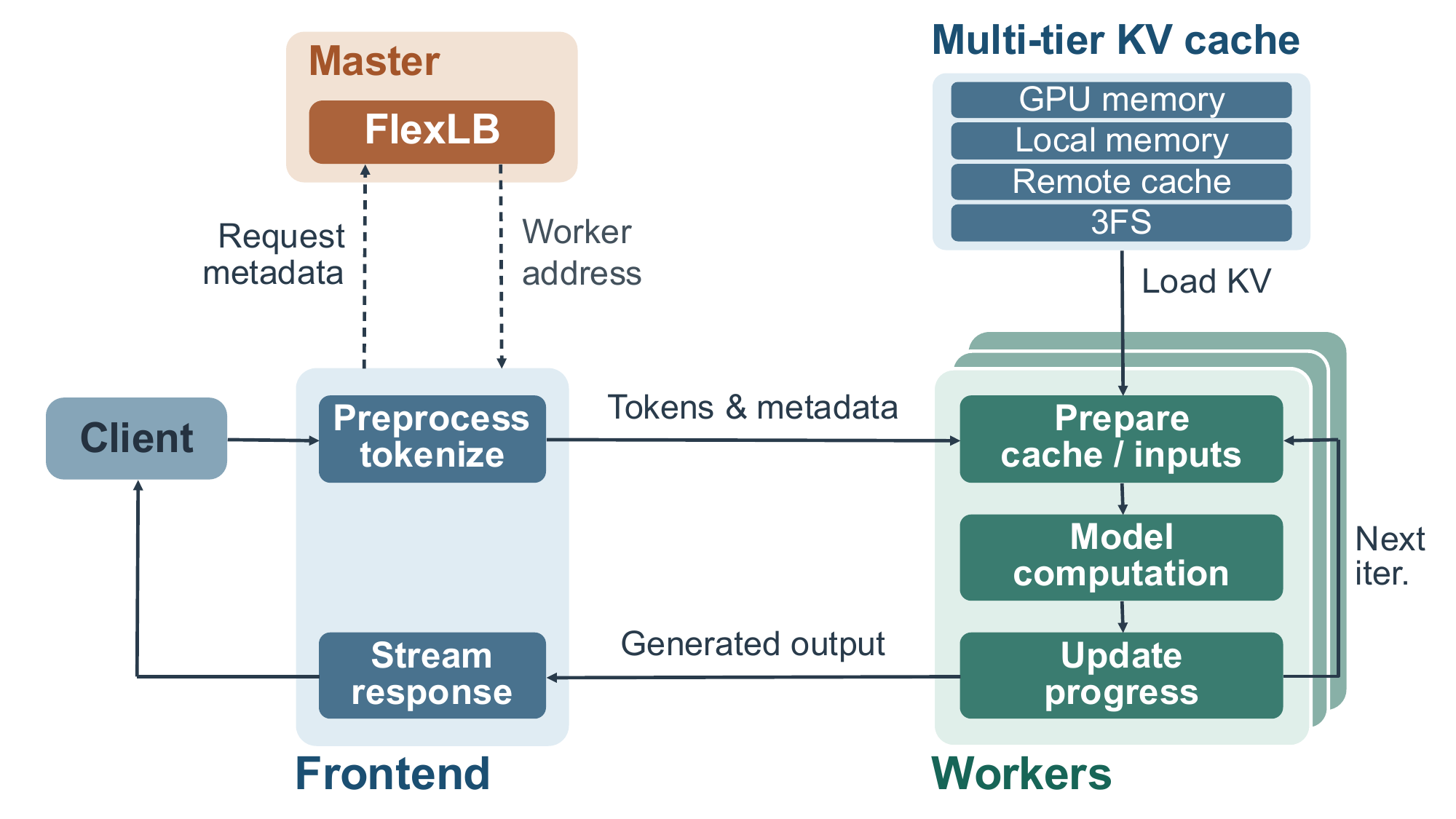}
    \caption{Request execution in \ourname{}.}
    \label{fig:architecture}
    \vspace{-0.18in}
\end{figure}

\autoref{fig:architecture} shows the main components of \ourname{}: a Frontend, a Master, a pool of inference workers, and a multi-tier KV cache subsystem.
The Frontend provides the application-facing service interface, while the Master coordinates request placement across the worker pool.
Workers provide model execution capacity and maintain the execution context of the requests assigned to them.
The cache subsystem supports the workers' state requirements through a storage hierarchy spanning worker-local memory and external storage.

For an autoregressive generation request, the Frontend first preprocesses the input and tokenizes the prompt.
It sends the prompt length and cache keys to the Master.
Flexible Load Balance (FlexLB), a component of the Master, uses cache-aware placement to select execution destinations, considering request characteristics, prefix-cache reuse, and worker load.
After receiving the worker addresses, the Frontend submits the tokenized input and generation parameters directly to the selected workers.

Each worker queues incoming requests and uses local scheduling to select the work for each model invocation.
The multi-tier KV cache spans GPU memory, local memory, remote cache, and 3FS.
Before processing the prompt, the runtime identifies reusable state for matching prefixes and loads it into GPU memory when needed.
It allocates space for new state and prepares token inputs, sequence lengths, and cache-block mappings for the scheduled requests.
The initial computation processes the uncached prompt tokens, establishing the state used by subsequent generation steps.

Generation then advances through successive model invocations.
After each invocation, the worker updates generated tokens and request progress and checks whether generation has completed.
Unfinished requests continue with their accumulated state; their execution inputs are rebuilt for the next invocation, with additional state storage allocated as needed.
As results become available, the Frontend converts the returned tokens into response content and streams it to the client until the request completes.

Within this organization, workers can use different execution configurations to meet model requirements and deployment objectives.
Configuration choices determine how model computation is distributed across devices and how the stages of request execution are assigned to workers.
For example, initial prompt processing (prefill) and iterative token generation (decode) can run on the same worker or on separate worker groups.
\autoref{tab:configurations} in \autoref{app:configurations} provides a fuller inventory of supported capabilities and configurations.

\section{Enabling Continuous Evolution}
\label{sec:evolvable}

\begin{figure}[t]
    \centering
    \includegraphics[width=0.95\linewidth]{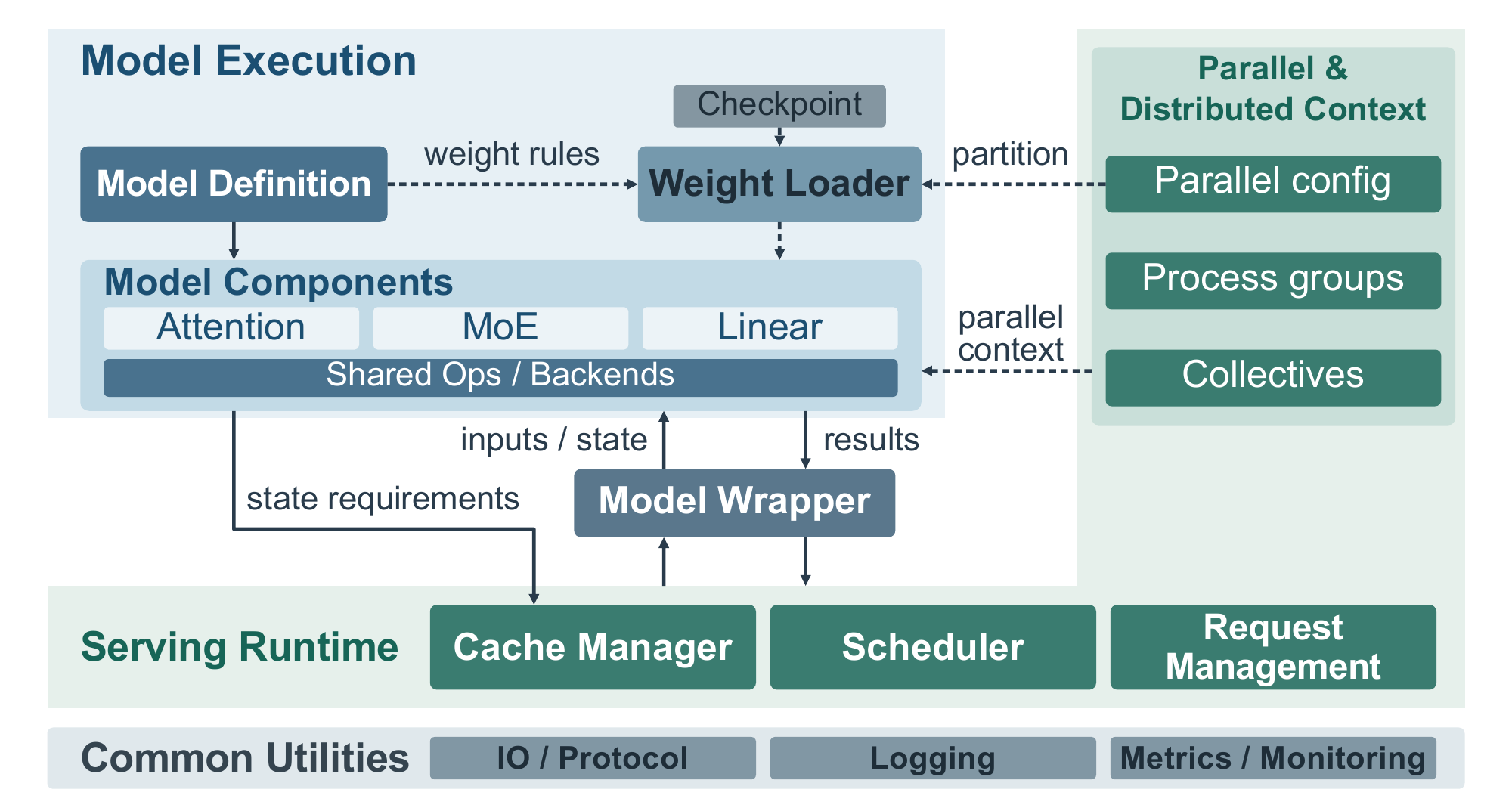}
    \caption{Model--serving separation of \ourname{}.}
    \label{fig:decouple}
    \vspace{-0.16in}
\end{figure}

Continuous model integration requires controlling how changes propagate through the serving system.
A new model can change computation and checkpoint layouts, as well as the state and distributed context needed for execution.
As shown in \autoref{fig:decouple}, \ourname{} places computation and parameter interpretation in model-side components, while request coordination and resource management remain in the shared serving runtime.
Explicit interfaces connect these responsibilities, allowing model-specific changes to remain local where possible and enabling different models to reuse common serving mechanisms.

\subsection{Localizing Model-Specific Changes}
\label{sec:model-extensibility}

Changes to model computation enter through Python/PyTorch definitions and components, keeping layer structure and operator choice separate from request coordination.
A model definition specifies the forward computation and uses the shared operator library (Ops) through these components.
Backend implementations register with component factories and expose support conditions on weights and execution configurations.
The factories use these conditions to select an implementation.
A hardware-specific kernel can therefore be added to the component library and reused by compatible models, while a new computation may require a dedicated component.
In both cases, the extension is organized around model computation, without introducing a separate request-serving path for each implementation.

Changes to checkpoint layouts are handled through model-supplied weight descriptions and conversion logic.
The descriptions map checkpoint tensors to internal parameters and specify required combinations or transformations.
The common loader applies these rules together with the selected parallel and numerical configuration to prepare executable weights.
This keeps parameter interpretation with the model while reusing the loading procedure.
The physical layout of prepared weights must match the selected kernel's requirements, even when tensor shapes are unchanged.

These model-side changes connect to serving through a common model invocation interface.
A shared model wrapper translates token and position inputs and sequence metadata into the inputs required by the Python model, including multimodal features when present.
For ordinary autoregressive generation, the model returns hidden states; the wrapper applies the output projection and prepares logits for the common sampling path.
When its invocation requirements fit this interface and its state requirements are supported by the runtime, changes to computation, weights, or operators can remain within the model and its associated components.
The serving runtime continues to invoke the model through this interface, so those changes do not require rewriting request scheduling or progress tracking.

\subsection{Reusing the Serving Runtime across Models}
\label{sec:runtime-extensibility}

Reusing the serving runtime requires more than a common model invocation: the runtime also needs to know what resources execution requires.
Models describe their state requirements and consume a shared parallel configuration defining their distributed context.
These descriptions and configurations let the runtime accommodate different model requirements through shared state-management and distributed-initialization mechanisms.

Persistent-state management illustrates how this reuse works.
The model supplies per-layer descriptors of state types and layouts, together with supported capacity, retention, and reuse policies.
Allocation and reuse depend on these policies as well as on state layout.
Configuration builders combine these descriptions with execution parameters to form state groups, and common cache-management interfaces handle their allocation and reclamation.
Allocated cache views are bound at model initialization; each invocation supplies sequence metadata and per-group block mappings that locate the current requests' state.
Models with different combinations of supported state types can therefore share allocation and lifecycle-management mechanisms through their descriptions.
Cache connectors use resource metadata from the same organization; models with layouts supported by a connector can also reuse its cache lookup and transfer path.

Parallel execution makes the distributed environment another point of reuse across models.
Common startup and communication utilities establish process groups from the shared parallel configuration and provide collective operations.
Model definitions and their components determine how these operations are combined with computation, allowing model-specific execution strategies to evolve while reusing common distributed initialization and communication support.

However, such reuse is limited to requirements that the runtime can already express and support.
A new state type or execution pattern may require changes to both the model's descriptions and the shared runtime implementation; exposing a new descriptor does not by itself provide the corresponding execution or resource-management support.
Such extensions incorporate the new requirements into shared serving mechanisms, preserving a common request lifecycle as supported models evolve.
\autoref{app:model-case-study} illustrates this process through DeepSeek-V4 integration, where new model computation and heterogeneous state requirements led to extensions on both sides of the interface.

\section{Preserving Production Reliability}
\label{sec:reliable}

Reusing a serving runtime across models also makes its reliability a shared concern.
Reliability must be maintained when requests acquire and release resources, while service instances fail or experience changing demand, and as shared components evolve.
\ourname{} addresses these concerns at three levels: state and resource protection within a worker, service availability and diagnosis across instances, and regression validation during evolution.

\subsection{State and Resource Protection}
\label{sec:state-protection}

Within a worker, request execution proceeds through scheduling, input and state preparation, model computation, and progress and output updates.
Unfinished requests continue through further iterations, while completion or failure initiates cleanup.
Autoregressive requests retain execution state across successive model invocations, so resource use extends beyond the work selected for a single iteration.
The KV-cache subsystem stores this persistent state, whose footprint can grow as a request progresses.

Before model execution, \ourname{} constrains resource acquisition through scheduling and allocation checks.
The worker-local scheduler bounds batch size and token work and, when configured, the number of requests with initialized KV resources.
The cache manager checks block availability when requests first acquire state and as that state grows.
Scheduling limits control current work; allocation checks enforce availability for the state that work requires.
Requests can wait for scheduling eligibility, while an attempted allocation that fails at initialization or during growth terminates the affected request with an allocation error.

During input and state preparation, the runtime checks data readiness.
Storage ownership keeps buffers live, while completion tracking determines when their contents are usable.
Cache transfers can be submitted asynchronously, so submission and publication are treated as distinct events.
Transfer contexts retain the storage they access, and destination entries become reusable only after a successful copy.
Before execution consumes loaded state, the request checks that its load has completed successfully.
These dependencies allow data movement to overlap with other work while preserving the ordering between production and consumption.

During model execution and progress updates, asynchronous operations can continue using a request's resources.
For decode-only batches in non-speculative execution, optional asynchronous output processing uses device events and explicit waits to preserve dependencies between computation, output transfer, and request-state updates.
When the request terminates, reclamation must preserve these completion dependencies.
Completion and error use a common cleanup path, with guards against repeated release.
Cleanup tracks outstanding output tasks and defers KV release until they complete.
A timeout can therefore stop response production while pending local work finishes safely.

State protection also extends beyond the originating request.
Prefix reuse lets later requests share retained state, and asynchronous transfers can hold blocks independently of active computation.
\ourname{} separately tracks references held by requests, cache connectors, and retained cache entries.
A block becomes available for unrelated allocation only after all references are released.
Evicting a prefix releases its retained-cache reference without overriding ownership by active requests or transfers.
Separating retention policy from active ownership allows cache-retention policies to change without invalidating ongoing work.
Keeping allocation, ownership tracking, and cleanup in the shared runtime carries these protections to models using these shared state-management mechanisms.

\subsection{Operational Reliability}
\label{sec:service-recovery}

\begin{figure}[t]
    \centering
    \includegraphics[width=0.95\linewidth]{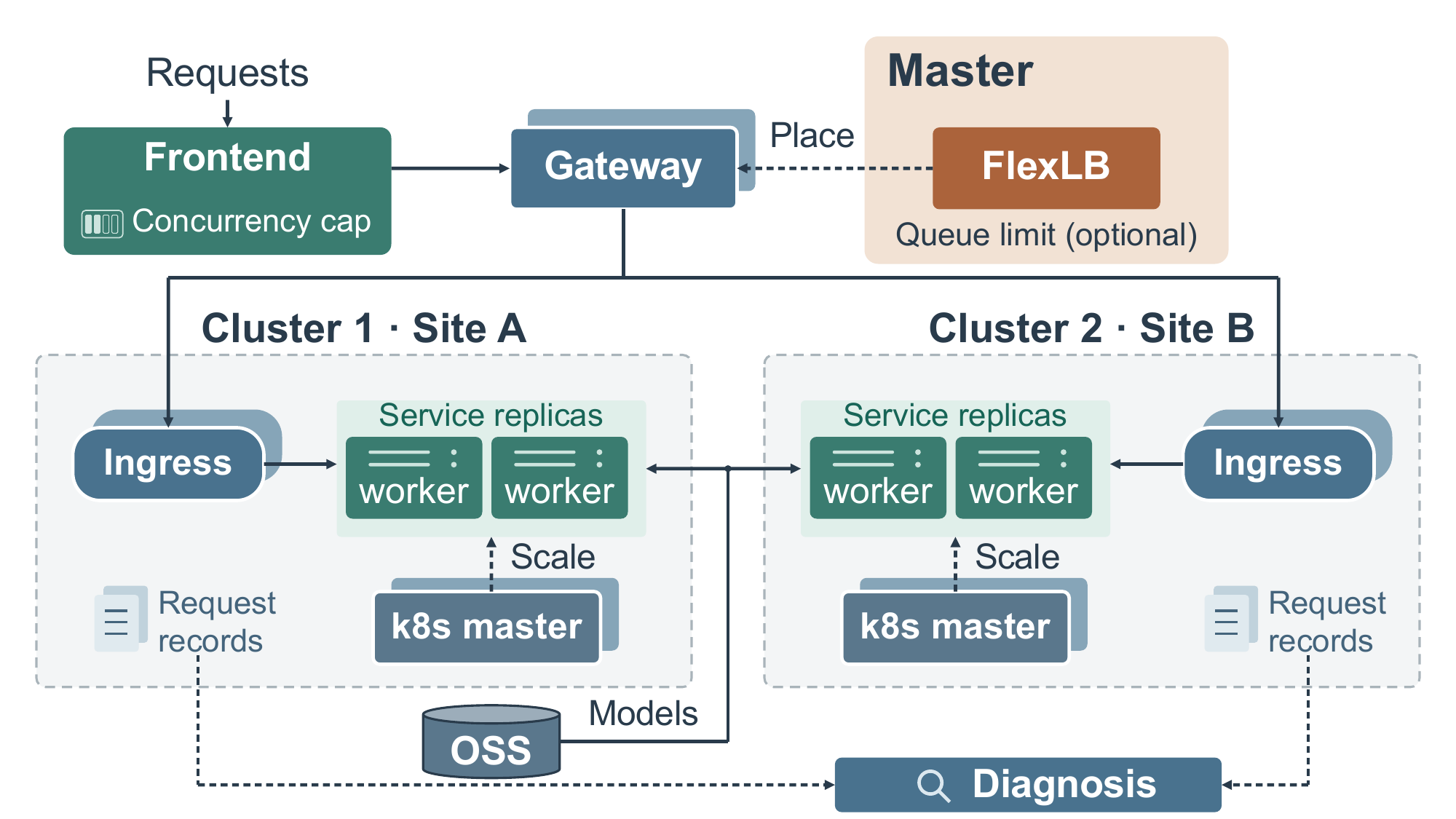}
    \caption{\ourname{} deployment across clusters and sites.}
    \vspace{-0.1in}
    \label{fig:deployment}
\end{figure}

Service availability depends on both worker capacity and the request-entry and coordination paths that make it reachable.
\autoref{fig:deployment} illustrates a typical deployment of \ourname{} supporting Alibaba's production services.
The deployment combines redundant instances of the same service with placement across multiple clusters and sites.
The Master uses FlexLB for request placement, while Gateway handles request forwarding; Ingress supports service routing within a cluster, and Kubernetes supports deployment control.
Worker replicas provide alternative capacity for subsequent requests after an instance failure, while placement across clusters and sites limits exposure to shared infrastructure failures.
Replication of Gateway and Ingress protects request access from individual instance failures, while Kubernetes control-plane replication protects deployment management.

To protect response latency as demand changes, Kubernetes manages worker replica scaling within a cluster.
Scaling decisions can use time to first token (TTFT) and time per output token (TPOT) alongside input and output tokens per minute (TPM), relating response latency to the token workload handled by the service.
These signals can guide adding workers when capacity is insufficient and reducing capacity under lower demand while maintaining latency objectives.

Admission limits and resource-aware placement complement scaling by constraining pressure on the available worker pool.
The Frontend enforces a configurable concurrency limit at request admission; when FlexLB queueing is enabled, its queue limit bounds requests awaiting placement.
FlexLB makes a worker selectable only after initializing its scheduling state and excludes workers reported as unavailable.
Within the available worker pool, resource-aware placement considers the work and state a request will require, with cache-aware policies also accounting for reusable prefixes.
Placement estimates can include work already assigned but not yet reflected in worker reports, reducing the risk of overcommitting workers between updates.
Worker-side allocation checks verify concrete resource availability during execution, complementing the estimates used for request placement.

Operational diagnosis supports recovery by locating request failures and latency degradation.
Request and error rates, response latency, queue length, and admission rejection expose failures and growing service pressure.
The Frontend forwards request identifiers and, when present, trace identifiers with generation requests to workers.
Access logs and execution-stage timings provide local evidence about slow or failed requests.
In production, distributed tracing uses the propagated identifiers to correlate these records along the execution timeline of a request.
This timeline lets operators locate delays and reported failures across the components that handle the request.

Output behavior provides a complementary view of service reliability.
An optional incremental monitor in the Frontend detects repeated token patterns.
Tool-call loop detection uses configured output-format markers.
When enabled, these checks report metrics and retain request context in logs.
Tool-call parsing diagnostics provide complementary evidence.
These signals make output anomalies observable alongside request failures and latency, supporting investigation even when requests complete successfully.

\subsection{Regression Validation during Evolution}
\label{sec:regression-validation}

Changes to a shared loader, operator, or serving component can affect configurations beyond the model that motivated them.
Regression validation must therefore cover the existing configurations that depend on the changed component.
The infrastructure combines component tests, end-to-end inference tests, and performance comparisons, with cases specifying model and execution settings and hardware profiles associating them with device types and counts.
For each shared-component change, the team selects regression configurations according to the affected models, numerical formats, parallel layouts, and serving paths.
Broader regression is performed before a production release to check interactions beyond the initial development configuration.

Validation checks behavior, performance, and resource safety under explicit conditions.
Behavioral comparisons specify request inputs, sampling settings, reference outputs, and tolerances; in validation mode, request and comparison errors propagate as failures.
Performance comparisons require comparable workload and execution settings across versions.
Resource-lifecycle tests check repeated allocation and release, cancellation and timeout states, and the ordering of completion visibility and reclamation, using expected block availability and release state.
Together, these checks evaluate observable results and the resource-management guarantees that changes to the shared runtime must preserve.

Validation also accounts for whether the intended comparisons were completed.
Completed results are interpreted with their case identities, hardware settings, and workload parameters.
The continuous integration system (CI) preserves reported failures.
Regression and production validation check report completeness and verify that the expected performance parameter points are present, while regression records also account for absent baselines and invalid measurements.
When the test infrastructure changes, coverage audits compare case identities, hardware assignments, and parameter settings.
This ties the coverage claimed for a shared-runtime change to completed comparisons on the configurations.

The release process includes resource-lifecycle checks.
A candidate that passes the required regression then enters a limited production rollout, where error rates, output abnormalities, latency, and resource usage are compared with the serving version.
Regressions pause further rollout and can trigger a return to the previous version.
This connects regression validation to the operational signals used to protect existing services as shared components evolve.

\section{Deployment}
\label{sec:deployment}

\ourname{} is deployed as part of Alibaba's shared inference infrastructure, serving production applications across domestic and international commerce, logistics, digital mapping, and local consumer services.
As of September 2026, \ourname{} serves 6,337 business units and supports $\sim$100 models, with deployments spanning multiple clusters and processing hundreds of trillions of tokens (input and output) per day.
Across successive engine versions, the system's development and production use span more than three years, and \ourname{} directly or indirectly serves hundreds of millions of users through the applications it supports.

Production workloads differ in inference task, input modality, prompt and output lengths, and traffic patterns.
They are served by models of different architectures and parameter scales running on heterogeneous accelerator platforms.
Parallel configurations and serving capabilities are selected to match the available hardware and each business's latency, throughput, and availability requirements.

\section{Evaluation}
\label{sec:evaluation}

In this section, we draw on data from model enablement and production operation at Alibaba to answer the following questions about \ourname{}:
\begin{enumerate}[leftmargin=2em]
    \item How does it accommodate new model requirements?
    \item How does it perform on production workloads?
    \item How reliable is it in long-term production use?
\end{enumerate}

\subsection{Model Enablement}
\label{sec:eval-enablement}

To assess how \ourname{} supports model evolution, we examine the code changes for initial functional support of DeepSeek-V4, GLM-5.2, MiniMax-M3, and Qwen3.5.
\autoref{tab:model-enablement-changes} reports the number of revised production-source lines, divided between the model and serving layers in \autoref{fig:decouple}.

\begin{table}[t]
    \centering
    \caption[Code changes for model enablement in \ourname{}.]{Code changes for model enablement in \ourname{}.\protect\footnotemark}
    \label{tab:model-enablement-changes}
    \begingroup
    \small
    \begin{tabularx}{\linewidth}{@{}l>{\raggedleft\arraybackslash}X>{\raggedleft\arraybackslash}X@{}}
        \toprule
        \textbf{Model} & \textbf{Model layer} & \textbf{Serving layer} \\
        \midrule
        DeepSeek-V4 & 19,749 (3.1671\%) & 5,207 (0.8350\%) \\
        GLM-5.2 & 423 (0.0678\%) & 38 (0.0061\%) \\
        MiniMax-M3 & 10,448 (1.6755\%) & 91 (0.0146\%) \\
        Qwen3.5 & 884 (0.1418\%) & 2 (0.0003\%) \\
        \bottomrule
    \end{tabularx}
    \endgroup
\end{table}
\footnotetext{Each percentage is the ratio of lines changed for that model in the corresponding layer to \ourname{}'s total production-source line count.}

For all four models, most changes are concentrated in the model layer and implement new computation, checkpoint interpretation, and supporting operators.
For example, MiniMax-M3 adds sparse attention and the associated support for quantized weights and operators, while DeepSeek-V4 introduces new attention and residual computations and native FP8/MXFP4 weight loading.
However, some models can be adapted using components already developed for earlier models.
For example, GLM-5.2 reuses GLM-5's MLA and expert components, while Qwen3.5 builds on Qwen3Next's linear-attention components.
These models thus require relatively few model-layer changes.

Serving-layer changes are relatively small compared with model-layer changes across all four adaptations.
GLM-5.2, MiniMax-M3, and Qwen3.5 each involve fewer than 100 changed serving-layer lines.
They reuse the existing request scheduler and cache manager by supplying model-specific state layouts and configurations to the common allocation, reuse, and reclamation mechanisms.
For DeepSeek-V4, we add the missing heterogeneous-state support by extending shared allocation, reuse, and cross-worker transfer mechanisms while retaining the common request-serving workflow.
Even with these extensions, its serving-layer changes remain substantially smaller than its model-layer changes.

Therefore, \ourname{}'s design supports continuous model integration while keeping the serving layer largely stable.
When new state or execution requirements demand runtime changes, the system extends shared mechanisms while preserving a common request-serving workflow across models.
To further demonstrate \ourname{}'s evolvability, we also report the model-enablement timelines in \autoref{app:model-enablement-timelines}, showing that these models reached production readiness in a short time after release.

\subsection{Production Serving Performance}
\label{sec:eval-performance}

\begin{figure*}[!t]
    \vspace{-0.04in}
    \centering
    \subfigure[Chat]{
        \begin{minipage}{0.305\textwidth}{
            \makebox[\linewidth][c]{\includegraphics[width=0.98\linewidth,trim=0 0 0 17bp,clip]{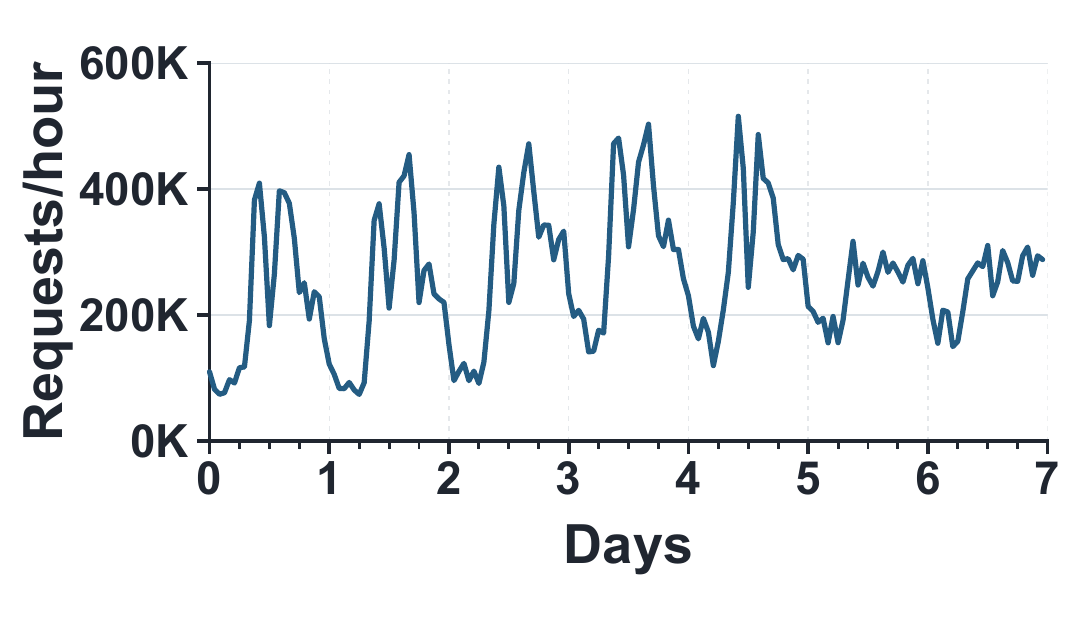}}
            \vspace{-0.22in}
        }\end{minipage}}\hfill
    \subfigure[Coding agent]{
        \begin{minipage}{0.305\textwidth}{
            \makebox[\linewidth][c]{\includegraphics[width=0.98\linewidth,trim=0 0 0 17bp,clip]{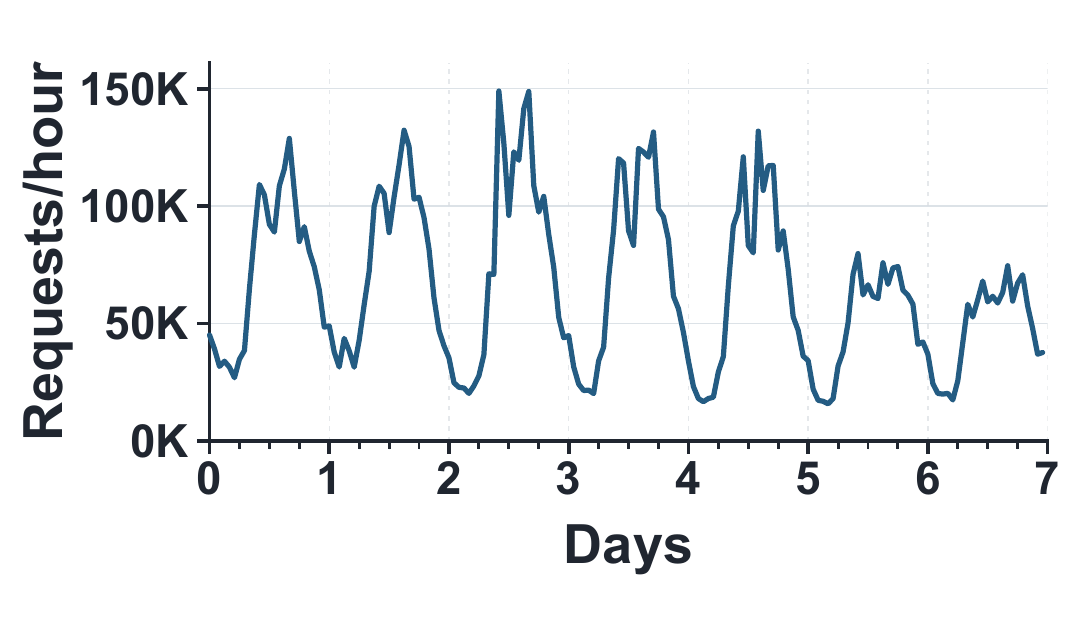}}
            \vspace{-0.22in}
        }\end{minipage}}\hfill
    \subfigure[Intelligent customer service]{
        \begin{minipage}{0.305\textwidth}{
            \makebox[\linewidth][c]{\includegraphics[width=0.98\linewidth,trim=0 0 0 17bp,clip]{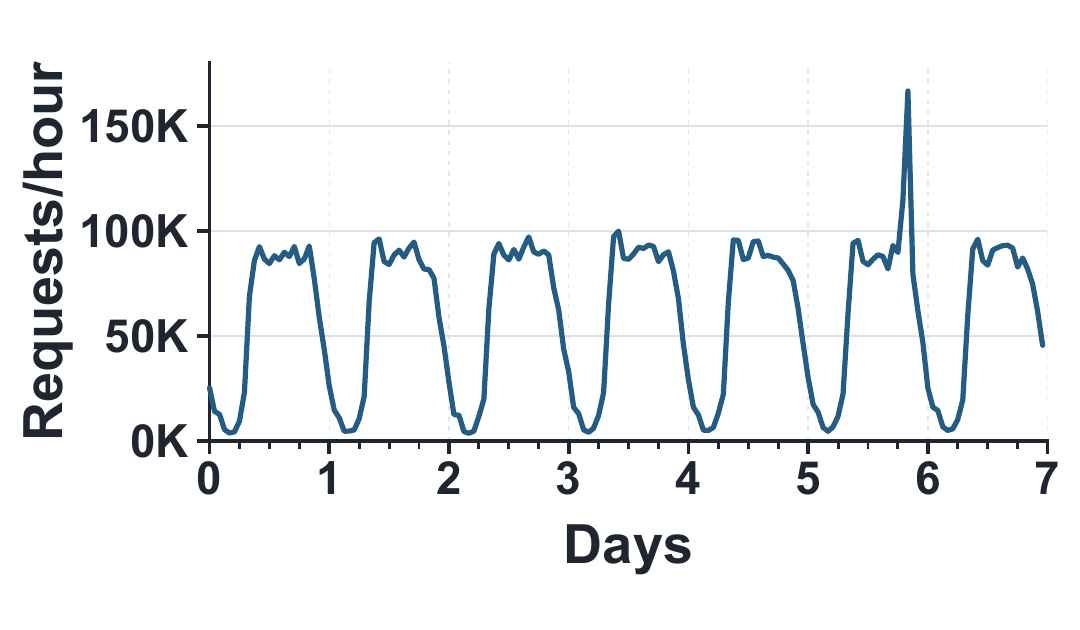}}
            \vspace{-0.22in}
        }\end{minipage}}%
    \vspace{-0.03in}
    \caption{Request volume over seven days.}
    \vspace{-0.01in}
    \label{fig:req}
\end{figure*}

\begin{figure*}[!t]
    \vspace{-0.04in}
    \centering
    \subfigure[Chat]{
        \begin{minipage}{0.305\textwidth}{
            \makebox[\linewidth][c]{\includegraphics[width=0.98\linewidth,trim=0 0 0 17bp,clip]{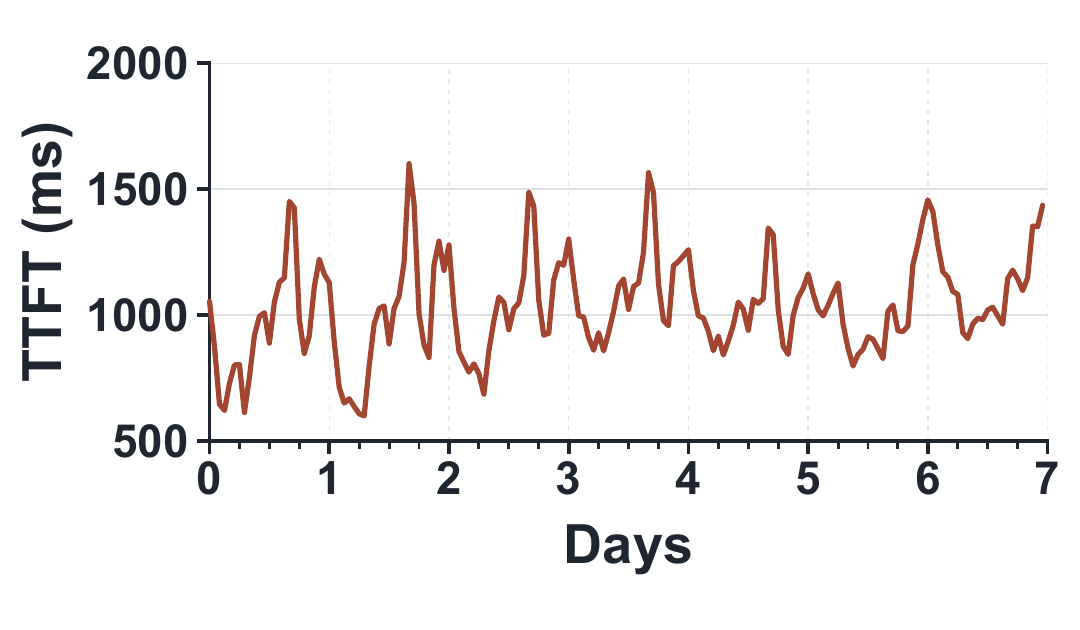}}
            \vspace{-0.22in}
        }\end{minipage}}\hfill
    \subfigure[Coding agent]{
        \begin{minipage}{0.305\textwidth}{
            \makebox[\linewidth][c]{\includegraphics[width=0.98\linewidth,trim=0 0 0 17bp,clip]{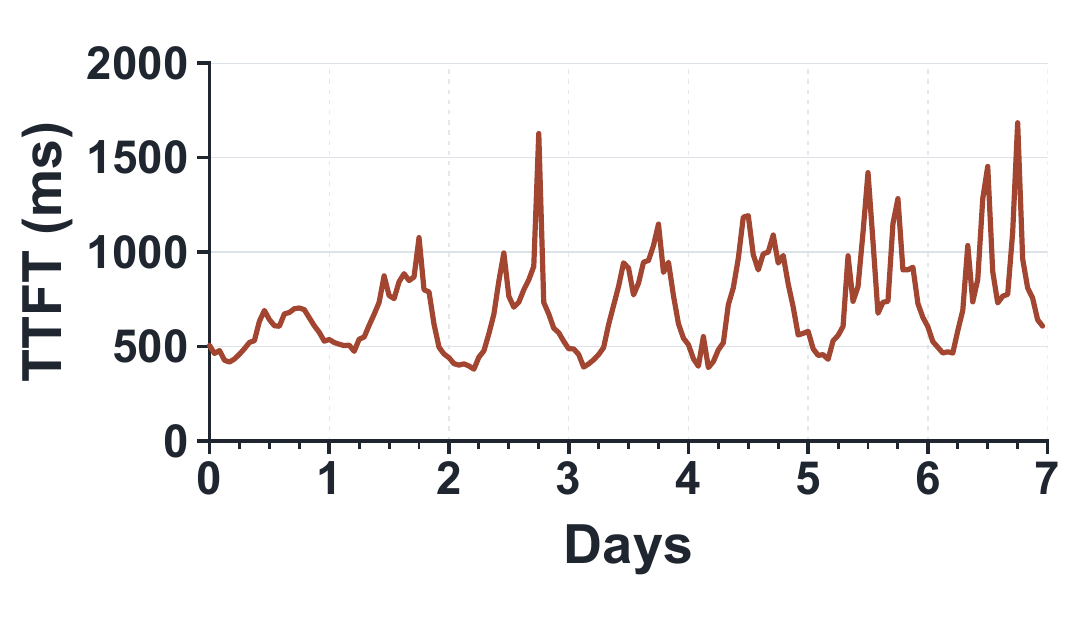}}
            \vspace{-0.22in}
        }\end{minipage}}\hfill
    \subfigure[Intelligent customer service]{
        \begin{minipage}{0.305\textwidth}{
            \makebox[\linewidth][c]{\includegraphics[width=0.98\linewidth,trim=0 0 0 17bp,clip]{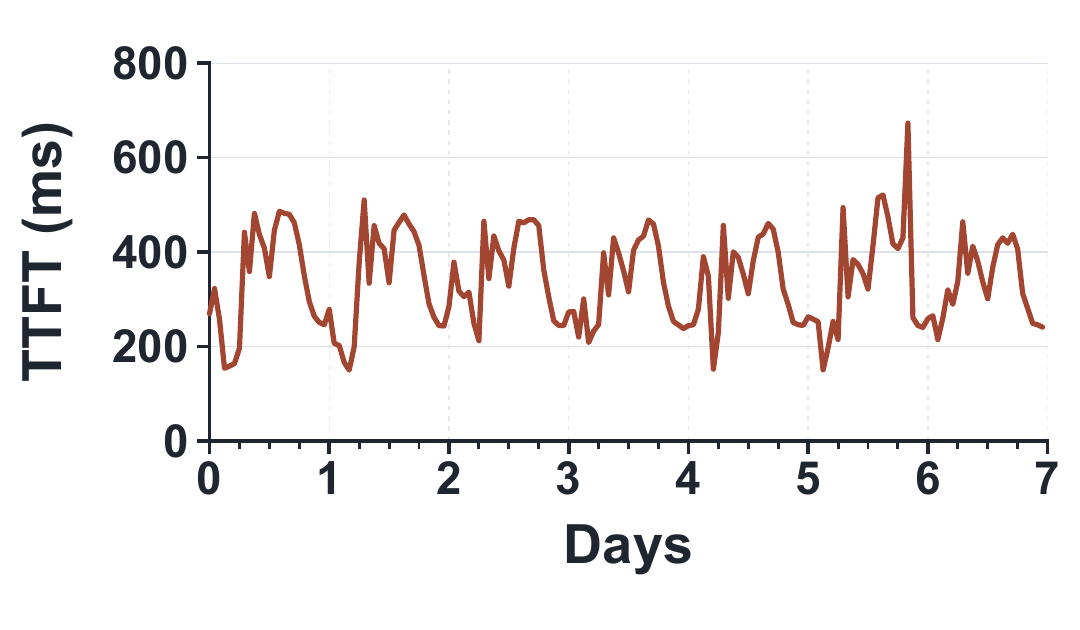}}
            \vspace{-0.22in}
        }\end{minipage}}%
    \vspace{-0.03in}
    \caption{Mean time to first token (TTFT) over seven days.}
    \vspace{-0.01in}
    \label{fig:ttft}
\end{figure*}

\begin{figure*}[!t]
    \vspace{-0.04in}
    \centering
    \subfigure[Chat]{
        \begin{minipage}{0.305\textwidth}{
            \makebox[\linewidth][c]{\includegraphics[width=0.98\linewidth,trim=0 0 0 17bp,clip]{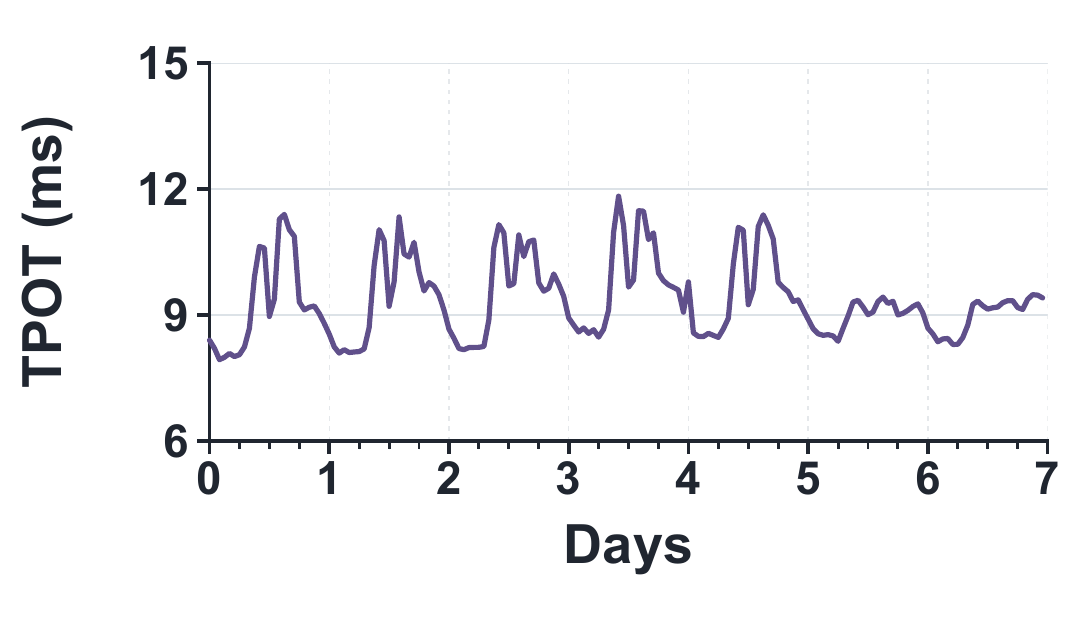}}
            \vspace{-0.22in}
        }\end{minipage}}\hfill
    \subfigure[Coding agent]{
        \begin{minipage}{0.305\textwidth}{
            \makebox[\linewidth][c]{\includegraphics[width=0.98\linewidth,trim=0 0 0 17bp,clip]{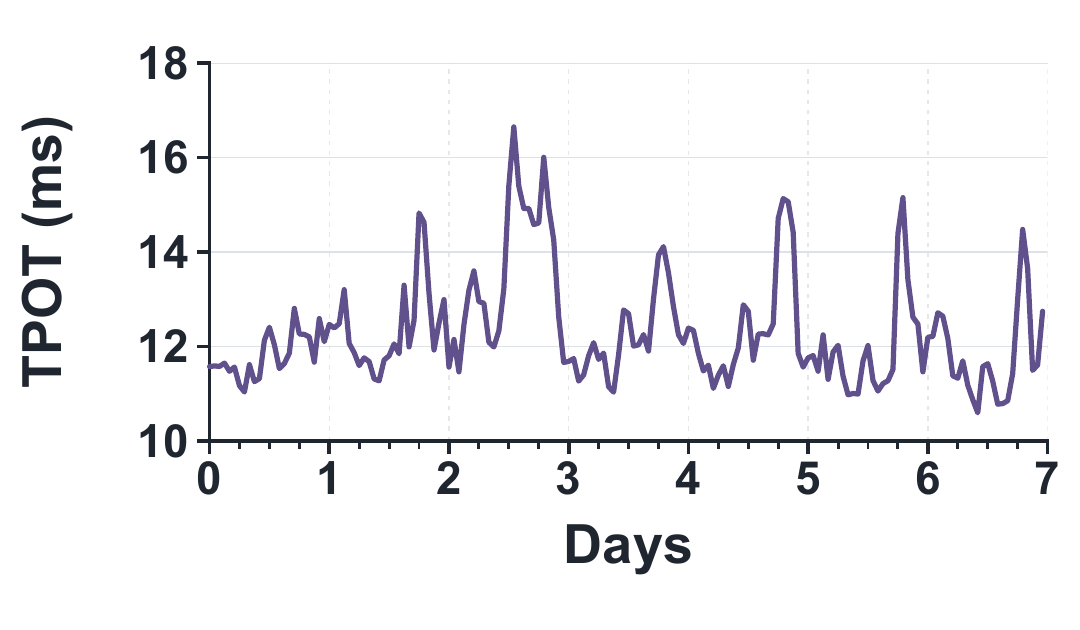}}
            \vspace{-0.22in}
        }\end{minipage}}\hfill
    \subfigure[Intelligent customer service]{
        \begin{minipage}{0.305\textwidth}{
            \makebox[\linewidth][c]{\includegraphics[width=0.98\linewidth,trim=0 0 0 17bp,clip]{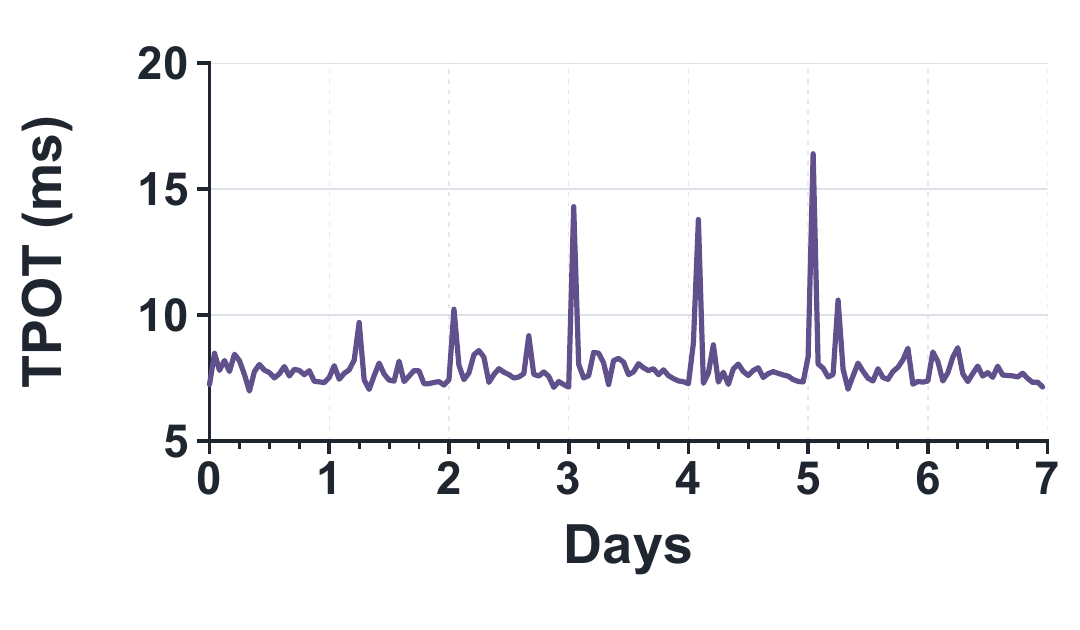}}
            \vspace{-0.22in}
        }\end{minipage}}%
    \vspace{-0.03in}
    \caption{Mean time per output token (TPOT) over seven days.}
    \vspace{-0.01in}
    \label{fig:tpot}
\end{figure*}

\begin{figure*}[!t]
    \vspace{-0.04in}
    \centering
    \subfigure[Chat]{
        \begin{minipage}{0.305\textwidth}{
            \makebox[\linewidth][c]{\includegraphics[width=0.98\linewidth,trim=0 0 0 8bp,clip]{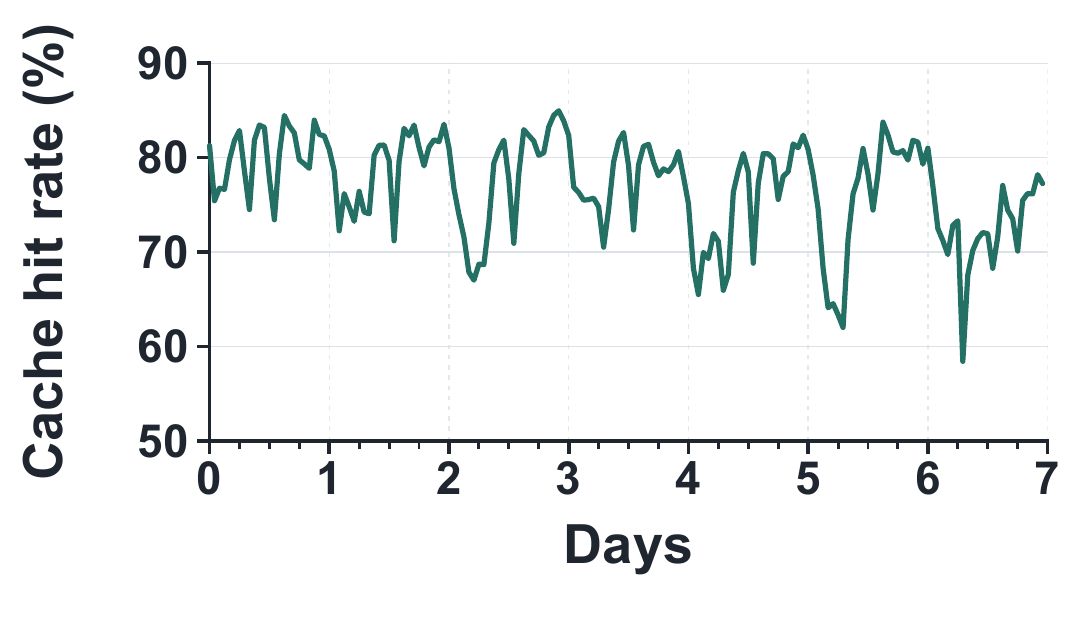}}
            \vspace{-0.22in}
        }\end{minipage}}\hfill
    \subfigure[Coding agent]{
        \begin{minipage}{0.305\textwidth}{
            \makebox[\linewidth][c]{\includegraphics[width=0.98\linewidth,trim=0 0 0 8bp,clip]{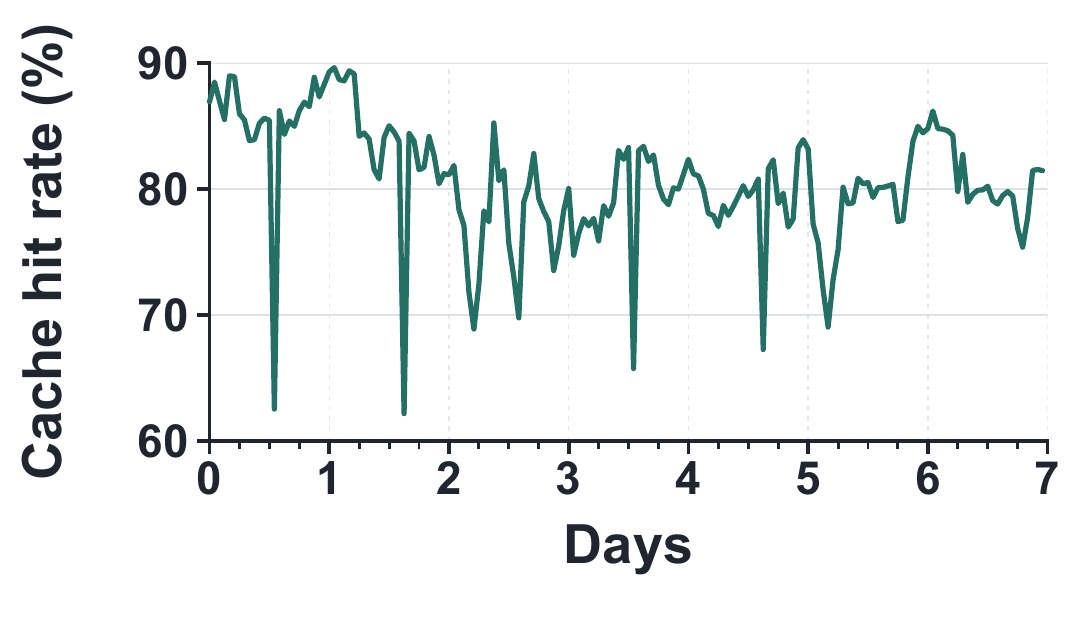}}
            \vspace{-0.22in}
        }\end{minipage}}\hfill
    \subfigure[Intelligent customer service]{
        \begin{minipage}{0.305\textwidth}{
            \makebox[\linewidth][c]{\includegraphics[width=0.98\linewidth,trim=0 0 0 8bp,clip]{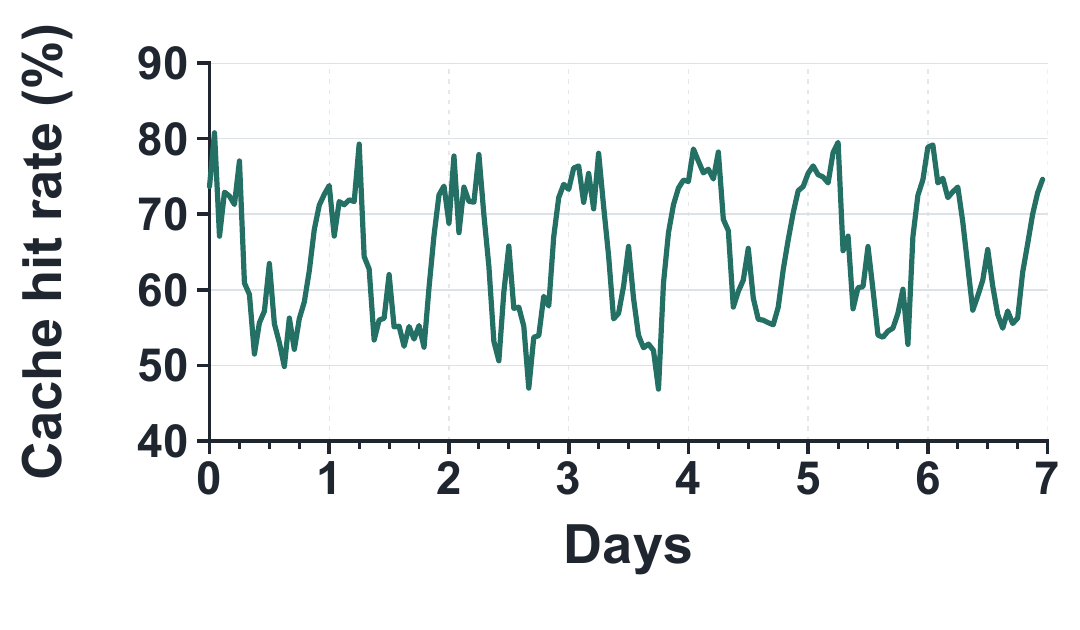}}
            \vspace{-0.22in}
        }\end{minipage}}%
    \vspace{-0.03in}
    \caption{Mean cache hit rate over seven days.}
    \vspace{-0.01in}
    \label{fig:cache}
\end{figure*}

We examine three production applications: chat, an internal coding agent, and intelligent customer service, each served by a different model in a separate production cluster.
Each set of measurements is collected from a single cluster over seven days to capture the recurring daily traffic cycle.
Figures~\ref{fig:req}--\ref{fig:cache} present hourly observations of request volume, mean time to first token (TTFT), mean time per output token (TPOT), and mean cache hit rate.

We observe recurring daily cycles in business demand, with request volumes rising during the daytime and falling late at night.
Across these changing loads, the reported TTFT remains below 2 seconds and TPOT below 20 milliseconds for all three applications, and the services also meet their respective request-level SLOs in production over the observation period.
We also observe cache reuse in all three applications, with hit rates of up to 80--90\%.
By reusing previously computed context, \ourname{} avoids repeated computation for requests with shared prefixes and reduces the computational load of production serving.

Although applications differ in characteristics and traffic patterns, \ourname{} allows each business to select a suitable model.
The shared runtime then provides responsive execution and effective cache reuse, meeting practical requirements with high performance.
Supporting these choices through a common runtime therefore combines evolvable model support with reliable production performance.

\subsection{Long-Term Service Reliability}
\label{sec:eval-reliability}

\begin{figure*}[!t]
    \vspace{0.00in}
    \centering
    \subfigure[Traffic]{
        \begin{minipage}{0.305\textwidth}{
            \makebox[\linewidth][c]{\includegraphics[width=0.98\linewidth]{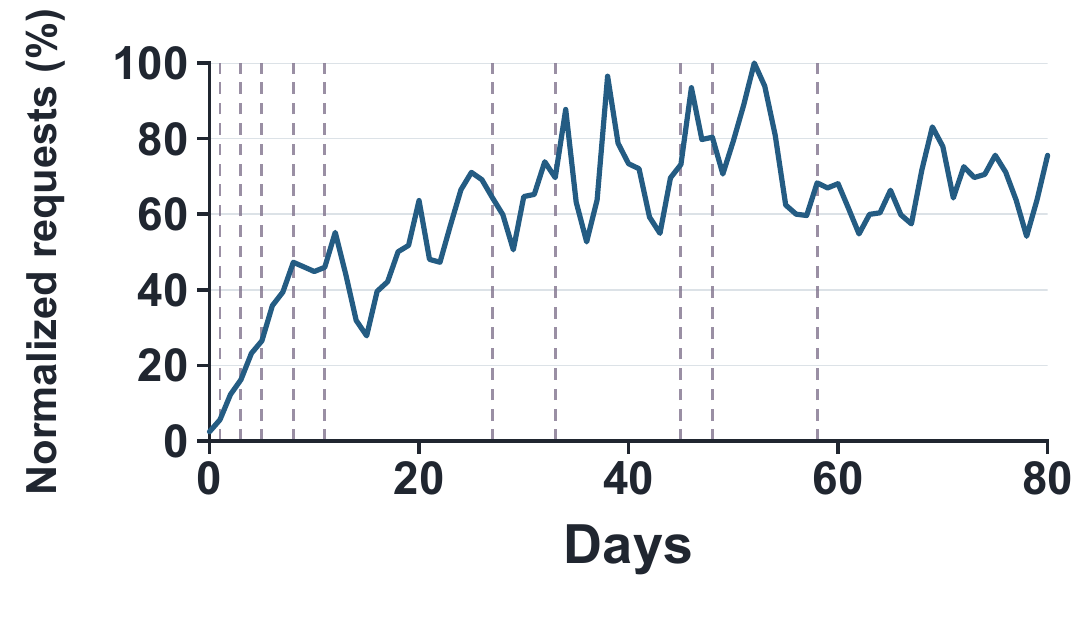}}
            \vspace{-0.22in}
        }\end{minipage}}\hfill
    \subfigure[Error rate]{
        \begin{minipage}{0.305\textwidth}{
            \makebox[\linewidth][c]{\includegraphics[width=0.98\linewidth]{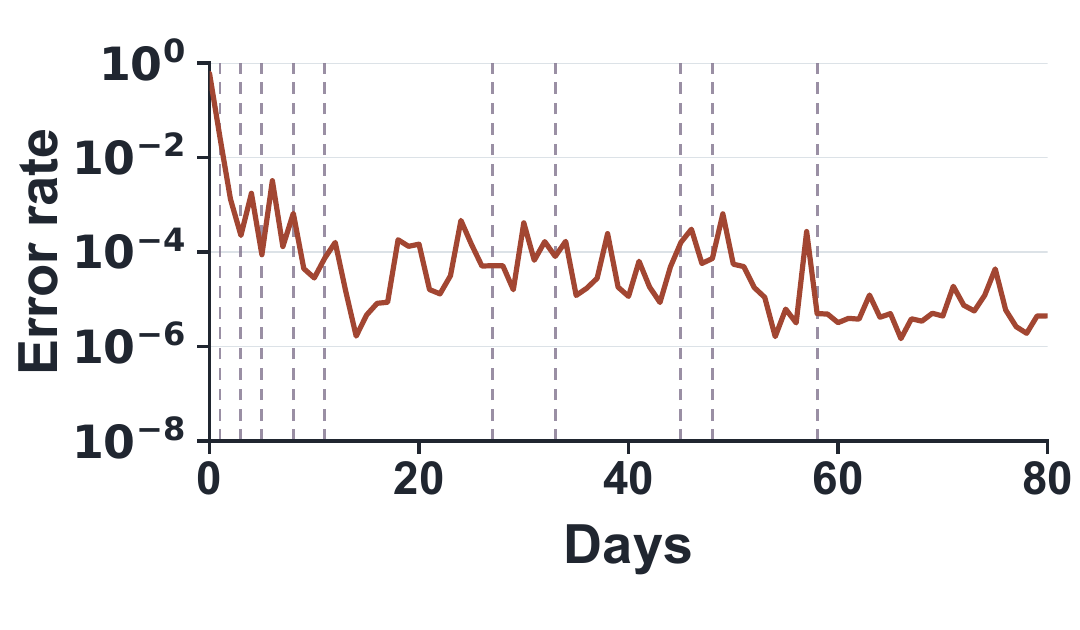}}
            \vspace{-0.22in}
        }\end{minipage}}\hfill
    \subfigure[Error breakdown]{
        \begin{minipage}{0.305\textwidth}{
            \makebox[\linewidth][c]{\includegraphics[width=0.98\linewidth]{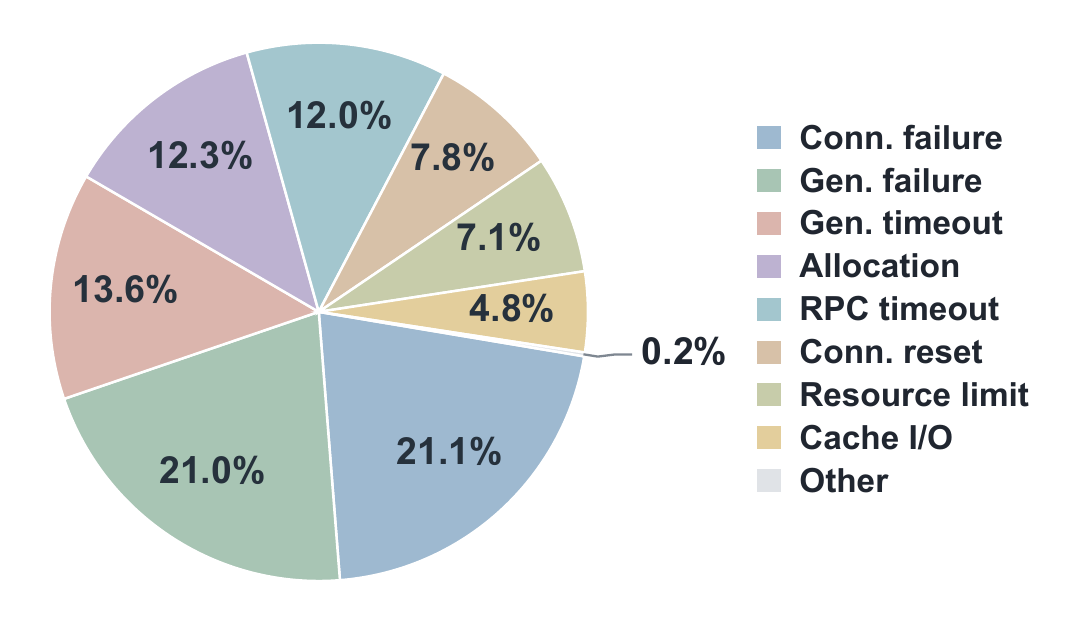}}
            \vspace{-0.22in}
        }\end{minipage}}%
    \vspace{-0.075in}
    \caption{Service reliability under frequent updates.}
    \vspace{-0.02in}
    \label{fig:service-reliability}
\end{figure*}

Figures~\ref{fig:service-reliability}(a) and~\ref{fig:service-reliability}(b) track a model-as-a-service (MaaS) offering over approximately 80 days, using measurements collected from a single production cluster.
The vertical dashed lines mark version updates, which typically involve bug fixes, performance tuning, inference optimizations, or improvements to diagnostic logging and system stability.
As the service gains broader adoption, its daily request volume grows while the request error rate declines across successive updates.
Some fixes markedly reduce errors associated with known bugs on the day of deployment, and the error rate remains below 0.01\% on nearly all days in the later part of the observation period.
These records illustrate \ourname{}'s ability to evolve while improving production reliability.

Figure~\ref{fig:service-reliability}(c) summarizes reported failure modes among failed requests over the same period.
Requests may fail because a worker cannot be located or reached, an established connection is reset, or a remote procedure call (RPC) times out; these cases account for 40.9\%.
Reported execution errors and exceeded generation deadlines form another major share (34.6\%), while failed memory or KV-cache allocations and reported overload or resource exhaustion contribute 19.4\%.
Cache loading and transfer can likewise encounter connection failures, timeouts, and data-transfer errors, accounting for a smaller 4.8\%.
The diversity of errors encountered in production requires complementary measures across the serving stack, combining bug fixes, resource protection, failure handling, and improved diagnostics to sustain reliable service as the system evolves.

\section{Lessons Learned}

\textbf{The long-term cost of owning an inference engine.}
Developing \ourname{} around Alibaba's requirements gives us flexibility in choosing which model, hardware, and serving capabilities to prioritize.
That flexibility comes with responsibility for their implementation, validation, optimization, and long-term maintenance.
Adopting an existing engine also requires local integration and production validation, so imperfections in an open-source framework are not sufficient grounds for starting an independent implementation.
The decision should therefore weigh the benefits of organization-specific evolution against the engineering and maintenance costs over the engine's lifetime.
When existing engines can accommodate an organization's requirements without extensive, recurring modifications, adopting and extending them may be more sustainable than maintaining an independent stack.

\textbf{Balancing execution efficiency with model-development cost.}
Our C++-first design prioritized execution efficiency, but model evolution exposed a recurring engineering cost.
Supporting new models required translating their Python/PyTorch reference implementations into C++, adding extra development overhead.
We therefore moved model execution to Python/PyTorch while retaining scheduling and KV-cache management in C++.
This adjustment reflects a practical trade-off: runtime overhead must be weighed against the recurring work of translating, validating, and maintaining rapidly changing model code, rather than applying the same implementation strategy to every component.

\textbf{Generalizing after concrete model adaptation.}
Some of our model adaptations introduced requirements that the shared framework did not yet express.
We have handled such changes by first establishing a working model adaptation, then generalizing its reusable capabilities into shared framework mechanisms.
Working through the adaptation makes the computation and state requirements concrete, helping distinguish model-specific details from functionality that belongs in a common abstraction.
The initial adaptation allows us to understand a requirement before committing to a general interface, but the follow-up consolidation remains part of its engineering cost.
Without that consolidation, locally successful adaptations risk accumulating as separate paths that must be maintained independently.

\textbf{Keeping cluster control understandable and recoverable.}
Our production experience highlights the importance of explicit component dependencies and failure boundaries in cluster control.
A request-placement policy should make clear which components it depends on and how it behaves when those components time out or become unavailable.
Coordination mechanisms should be designed to contain such failures, allowing operators to trace automated decisions and identify the affected part of the system.
Cluster control should also preserve a way for operators to redirect traffic manually when automated decisions become ineffective or difficult to diagnose.
This makes operational control an explicit design requirement: operators need a way to intervene even before the underlying fault has been fully diagnosed.

\textbf{Balancing request admission with service quality.}
We encountered resource overload when the Master admitted work without rate limiting.
This experience showed that admitting more requests can compromise the service delivered to requests already in progress.
We therefore treat admission limits as part of production reliability, rather than relying solely on redundancy and capacity expansion to absorb excess demand.
Rejecting requests has an immediate service cost, but continuing to accept work beyond available capacity can spread that cost across a larger set of requests through contention, queueing, and timeouts.
The appropriate balance depends on workload characteristics and service objectives: the aim is not to minimize rejection in isolation, but to preserve useful service under constrained capacity.

\textbf{Developing inference engines in the era of coding agents.}
LLM-based coding agents are changing the engineering effort required to support new models and optimize their execution in inference engines.
With reference implementations and defined correctness comparisons, agents can quickly implement model support and help bring a newly released model to correctness-ready.
However, reaching performance-ready requires assessing the model's performance potential on the target hardware and identifying the most promising optimization directions, both of which depend on understanding how model computation, data movement, and communication interact under the intended workload and deployment configuration.
At present, experienced engineers remain central to this assessment, while agents can assist with implementing and evaluating the chosen optimizations.
As initial adaptation becomes easier, performance analysis and optimization design account for a greater share of the expertise required to develop an inference engine.

\textbf{\ourname{}'s general applicability.}
Alibaba's scale amplifies the challenges addressed by \ourname{}, but these challenges do not arise from traffic volume alone.
For an organization serving a small, stable set of models on uniform hardware, a deployment tailored to those configurations may be sufficient.
However, as its model portfolio expands and release cycles accelerate, model support becomes a recurring integration task rather than a one-time implementation.
When new capabilities require changes to shared serving code, each integration also introduces validation and maintenance work for existing services.
Adding accelerator capacity does not remove this engineering burden, which can become significant well before an organization reaches Alibaba's deployment scale.
We therefore believe that the design space explored by \ourname{}---accommodating rapid model evolution while preserving a reusable and reliable serving foundation---will become increasingly relevant to organizations operating diverse, continuously changing model portfolios, rather than remain a concern unique to hyperscale deployments.

\section{Related Work}

LLM inference engines have developed complementary mechanisms for organizing autoregressive computation and managing its execution state.
Orca \cite{yu2022orca} introduces iteration-level scheduling to coordinate requests at the granularity of individual model iterations.
vLLM's PagedAttention \cite{vllm} reduces KV-cache fragmentation and redundant copying while supporting memory sharing across requests.
SGLang \cite{sglang} combines an interface for structured language model programs with runtime mechanisms such as prefix KV-cache reuse.
TensorRT-LLM \cite{tensorrt-llm} provides optimized execution through specialized kernels, numerical formats, and distributed execution support.
These systems provide important architectural and implementation precedents for combining model execution, request scheduling, and state management within an inference engine.
Their relevance extends beyond individual optimizations to the broader organization of the serving stack.

Complementary research improves model execution efficiency by reducing computation, data movement, and sequential decoding overhead.
Hardware-aware attention implementations reduce memory traffic and improve the utilization of GPU execution resources \cite{fa1,fa2, fa3,ye2025flashinfer}.
Low-bit quantization reduces the storage and bandwidth requirements of model weights and, in some methods, activations \cite{tseng2024quip,ashkboos2024quarot,liu2025spinquant}.
Speculative and multi-token decoding approaches generate candidate continuations and verify them in parallel, amortizing target-model execution over multiple output tokens \cite{cai2024medusa,li2024eagle,li2026eagle}.
These directions address different sources of inference overhead and introduce execution choices whose applicability depends on model structure, numerical representation, and hardware capabilities.
Together, they expand the set of execution mechanisms that inference engines must accommodate as models and accelerator platforms evolve.

Beyond individual model execution, serving systems optimize how requests share computation, state, and distributed resources.
Iteration-level scheduling and chunked-prefill batching organize prompt processing and token generation to balance throughput with request latency \cite{agrawal2024taming,holmes2024deepspeed}.
Other work reduces repeated prompt computation and state-loading overhead through prefix or segment-level reuse, selective recomputation, and compressed KV-cache transfer \cite{sglang, gim2024prompt,yao2025cacheblend,liu2024cachegen}.
Prefill--decode disaggregation separates execution phases to enable resource allocation and placement suited to their different demands \cite{qin2026mooncake,patel2024splitwise,zhong2024distserve}.
Runtime request migration and elastic parallelism further adapt execution placement and resource assignments to workload variation \cite{sun2024llumnix,wu2024loongserve}.
These approaches address complementary aspects of serving efficiency, from batch formation to cluster-wide resource organization.
Our work complements them by documenting how a shared inference engine incorporates evolving execution and serving capabilities, together with the architectural trade-offs and operational lessons arising from long-term production deployment.

\section{Conclusion}

We present \ourname{}, an open-source inference engine developed for continuous evolution and reliable production serving at Alibaba.
Our experience highlights the need to revisit the boundary between model execution and the serving runtime as computation and state requirements evolve.
Over more than three years of development, \ourname{} has grown from an exploratory system into shared production infrastructure serving diverse Alibaba businesses.
Its current architecture has continued to accommodate new models and serving capabilities without another fundamental redesign.
We share our engineering experience and lessons learned from this evolution.
As model portfolios expand and release cycles accelerate, we hope these lessons and the idea of separating model execution from the serving runtime will inspire practitioners building inference infrastructure around their own workloads and operational constraints.

% \newpage
% \clearpage

\bibliography{reference}
\bibliographystyle{mlsys2025}

% \section*{AI Usage Disclosure}

% The authors developed the core ideas and formulated the lessons learned from their experience designing, deploying, and operating \ourname{}.
% Generative AI tools, including Qwen, ChatGPT, Qoder, and Codex, were used to assist with drafting and revising the manuscript, polishing its language, and preparing figures and plotting code based on technical material and data provided by the authors.
% This assistance also included editorial feedback and cross-checking technical descriptions against the implementation.
% All claims and AI-generated outputs were reviewed and verified by the authors, who take full responsibility for all content in the paper.

\newpage
\clearpage
\appendix

\begin{table*}[!t]
    \vspace{-0.025in}
    \centering
    \small
    \caption{Representative configuration dimensions and capabilities of \ourname{}.}
    \label{tab:configurations}
    \begin{tabular}{p{0.22\textwidth}p{0.7\textwidth}}
        \toprule
        \textbf{Dimension} & \textbf{Representative capabilities and configurations} \\
        \midrule

        \multicolumn{2}{l}{\textit{Model-Specific Capabilities and Requirements}} \\
        \addlinespace[2pt]
        Workloads and modalities
            & Generation, embedding, reranking, and classification across text, vision--language, and audio inputs \\
        Model architectures
            & Dense and MoE models; MHA/MLA, compressed sparse attention, sliding-window attention, and recurrent/attention hybrids \\
        Numerical formats
            & FP16/BF16 and representative INT8/INT4, FP8, and FP4 formats for weights, activations, and KV state \\
        Speculative decoding
            & MTP, EAGLE/EAGLE-3, and DSpark block-draft decoding \\
        \midrule
        \multicolumn{2}{l}{\textit{Shared Runtime and Serving Mechanisms}} \\
        \addlinespace[2pt]
        Hardware backends
            & NVIDIA GPUs (CUDA) and AMD GPUs (ROCm) \\
        Shared operators
            & Reusable attention, MoE, and quantized-computation operators \\
        Parallel execution
            & Tensor, pipeline, data, and expert parallelism; prefill-context and FFN-sequence parallelism; multi-node execution \\
        Execution acceleration
            & CUDA Graphs; asynchronous output processing; expert load balancing and communication--computation overlap \\
        Service interfaces
            & Streaming OpenAI-compatible chat API; batch chat extension; native task APIs \\
        Generation controls
            & Sampling and beam search; JSON/schema, regex, EBNF, and structural-tag constraints \\
        Request scheduling
            & Cache-aware load balancing; worker-local continuous batching and chunked prefill \\
        Inference-state management
            & Grouped state pools; paged allocation; prefix reuse with prefix-tree indexing \\
        Cache storage
            & Device memory, host memory, local disk, and remote/distributed cache, including 3FS \\
        Deployment topology
            & Fused or disaggregated computation; co-located or separated service components \\
        Startup acceleration
            & Parallel and GPUDirect weight loading; reuse of JIT compilation artifacts \\
        \bottomrule
    \end{tabular}
    \vspace{-0.08in}
\end{table*}

\section{Supported Configurations}
\label{app:configurations}

\autoref{tab:configurations} organizes representative capabilities of \ourname{} from two perspectives: model-specific capabilities and requirements, and shared runtime and serving mechanisms. 
Together, they summarize the range of model requirements accommodated by RTP-LLM and the common mechanisms it provides for production serving.

\section{A Model-Integration Case Study}
\label{app:model-case-study}

We use DeepSeek-V4 \cite{deepseekv4} as a case study to illustrate how \ourname{} accommodates new models and their associated execution and serving requirements.

DeepSeek-V4 required more than extending an existing model configuration. 
Its layer-dependent combination of compressed sparse attention, heavily compressed attention, and sliding-window attention, together with mHC residual transformations and revised MoE routing, could not be directly expressed by the existing DeepSeek execution path. 
These components also introduced new checkpoint layouts and quantization requirements. 
Beyond computation, individual layers required multiple persistent states with different layouts, growth patterns, and reuse conditions. 
Supporting the model therefore required both new model implementations and extensions to shared state-management and execution mechanisms.

\ourname{} accommodated these changes at distinct extension points. 
A dedicated Python model implementation composed the new attention, residual, and MoE computations while integrating with the existing model initialization and invocation framework. 
Per-layer weight descriptions map V4's parameters into the common loading pipeline, while V4-specific conversion strategies handle its FP8 and MXFP4 weights, UE8M0 scales, and execution-specific layouts.
For example, its FP8 loading strategy preserves checkpoint-provided scales rather than re-deriving them through requantization; this behavior is restricted to V4 weights. 
On the serving side, heterogeneous cache groups represent compressed KV entries, indexer state, and sliding-window state, with corresponding allocation and reuse policies. 
The integration also adapted per-group block mappings and cross-worker cache transfer to V4's heterogeneous states, and adjusted MTP hidden-state buffers and CUDA Graph input buffers to accommodate its mHC-expanded representations.

The integration thus involved substantial model-specific code and meaningful runtime changes, but retained the common request-serving organization rather than introducing a separate serving architecture for DeepSeek-V4. 
Requests continued to use the existing request queues, batch-scheduling framework, and output-dispatch path, with V4-specific computation invoked through the common model interface.
New computation and parameter interpretation were implemented in the model and associated execution and loading modules, while cross-cutting state requirements were accommodated through extensions to shared mechanisms. 
This case illustrates that evolvability is not about keeping the runtime fixed, but about ensuring that model-driven changes can be absorbed through explicit extension points without fragmenting the shared serving path.

\section{Model Enablement Timelines}
\label{app:model-enablement-timelines}

We report readiness timelines for the four models examined in \autoref{sec:eval-enablement}.
\autoref{tab:model-enablement-timeline} summarizes their enablement timelines using the readiness milestones from \autoref{sec:model-readiness}, where we distinguish the time spent establishing model correctness, optimizing execution for production, and carrying out business integration and canary testing.
Dates indicate when each readiness milestone was reached.\footnote{
Timelines depend on team size and the engineering time allocated to each model, and are not intended as controlled comparisons of adaptation difficulty or engineering effort across models.}

\begin{table*}[!t]
    \vspace{-0.025in}
    \centering
    \caption{Model-enablement timelines for representative models in \ourname{}.}
    \label{tab:model-enablement-timeline}
    \begingroup
    \small
    \setlength{\tabcolsep}{2.5pt}
    \renewcommand{\arraystretch}{1.15}
    \begin{tabularx}{0.92\textwidth}{@{}>{\centering\arraybackslash}Xc*{3}{>{\centering\arraybackslash}p{0.17\textwidth}}>{\centering\arraybackslash}p{0.16\textwidth}@{}}
        \toprule
        \multirow{2}{*}{\textbf{Model}}
            & \multirow{2}{*}{\textbf{Available}}
            & \multirow{2}{*}{\textbf{Correctness-ready}}
            & \multirow{2}{*}{\textbf{Performance-ready}}
            & \textbf{Integration/}
            & \multirow{2}{*}{\textbf{Production-ready}} \\
            & & & & \textbf{Canary-ready} & \\
        \midrule
        DeepSeek-V4
            & 2026-04-24 & 2026-04-30
            & 2026-05-20 & 2026-05-21
            & 2026-05-22 \\
        GLM-5.2
            & 2026-06-17 & 2026-06-18
            & 2026-06-28 & 2026-06-30
            & 2026-07-01 \\
        MiniMax-M3
            & 2026-06-12 & 2026-06-15
            & 2026-06-26 & 2026-07-03
            & 2026-07-06 \\
        Qwen3.5
            & 2026-02-16 & 2026-02-18
            & 2026-03-02 & 2026-03-09
            & 2026-03-11 \\
        \bottomrule
    \end{tabularx}
    \endgroup
    \vspace{-0.12in}
\end{table*}

The table shows that reaching correctness-ready takes one to three days for GLM-5.2, MiniMax-M3, and Qwen3.5, and about one week for DeepSeek-V4.
By comparison, performance optimization accounts for the longest interval, at approximately three weeks for DeepSeek-V4 and two weeks for the other three models.
Since our goal is to match and improve upon community implementations while meeting production SLOs, this stage requires bottleneck analysis and optimization, performance benchmarking under comparable conditions, and checks for correctness and performance regressions. 
The results often prompt further tuning, so reaching performance-ready requires multiple rounds of optimization and testing.

The table also shows shorter integration/canary intervals, ranging from one day to about one week.
This is partly because integration with production service interfaces can often proceed alongside correctness validation and performance optimization.
Canary testing can also be interleaved with optimization, with successive optimized versions receiving limited production traffic.
After these checks and canary validation, the models become production-ready for Alibaba's internal applications and customer-facing services.

In general, the full process usually takes 14--28 days from community release.
These timelines illustrate \ourname{}'s ability to evolve with new model releases while completing the optimization and validation required for reliable production service.

\end{document}